\documentclass[american,aps,pra,reprint,superscriptaddress,floatfix,longbibliography]{revtex4-2}

\usepackage{graphicx,graphics,epsfig}   
\usepackage{dcolumn}    
\usepackage{bm}         
\usepackage{amsmath}    
\usepackage{verbatim}   
\usepackage{color}      
\usepackage{subfigure}  
\usepackage{times,natbib}
\usepackage{amsmath,amsfonts,amssymb,graphics,graphics,color,times}
\usepackage{hyperref}
\usepackage{mathtools}
\usepackage[T1]{fontenc} 

\usepackage{xcolor} 
\usepackage{url}

\usepackage{latexsym}
\usepackage{amsmath}
\usepackage{amssymb}
\usepackage{amsfonts}
\usepackage{amsthm}
\usepackage{mathrsfs}
\usepackage{color,verbatim,graphics}
\usepackage{psfrag}
\DeclareMathAlphabet{\mathrsfs}{U}{rsfs}{m}{n}
\DeclareMathAlphabet{\mathpzc}{OT1}{pzc}{m}{it}
\DeclareMathAlphabet{\matheus}{U}{eus}{m}{n}
\DeclareMathAlphabet{\mathbbold}{U}{bbold}{m}{n}

\newtheorem{theorem}{Theorem}[section]

\newtheorem{corollary}[theorem]{Corollary}
\newtheorem{lemma}[theorem]{Lemma}
\newtheorem{definition}[theorem]{Definition}

\theoremstyle{plain}

\def\one{\leavevmode\hbox{\small1\normalsize\kern-.33em1}}

\newcommand{\CC}{\mathbb{C}}

\renewcommand{\qed}{\ensuremath{\hfill \blacksquare}}

\def\one{\leavevmode\hbox{\small1\normalsize\kern-.33em1}}

\newcommand{\ba}{\begin{eqnarray}}
\newcommand{\ea}{\end{eqnarray}}
\newcommand{\ban}{\begin{eqnarray*}}
\newcommand{\ean}{\end{eqnarray*}}
\newcommand{\Tr}{\operatorname{Tr}}

\newcommand{\ket}[1]{|#1\rangle}
\newcommand{\bra}[1]{\langle#1|}

\newcommand{\Man}[1]{\left\|#1\right\|_1}
\newcommand{\Mat}[2]{\mathcal M_{#1, #2}}
\newcommand{\SM}[1]{\Man{#1}}
\newcommand{\Id}[1]{\mathbb I^{#1}}
\newcommand{\W}[2]{\mathcal W_{#1, #2}}
\newcommand{\Perm}[1]{\mathcal P_{#1}}
\newcommand{\transp}[1]{#1^\top}

\newcommand{\sign}{\text{sgn}}

\begin{document}

\title{Certification of qubits in the prepare-and-measure scenario with large input alphabet and connections with the Grothendieck constant}

\author{P\'eter Divi\'anszky}
\affiliation{MTA Atomki Lend\"ulet Quantum Correlations Research Group, Institute for Nuclear Research, P.O. Box 51, H-4001 Debrecen, Hungary} 

\author{Istv\'an M\'arton}
\affiliation{MTA Atomki Lend\"ulet Quantum Correlations Research Group, Institute for Nuclear Research, P.O. Box 51, H-4001 Debrecen, Hungary} 

\author{Erika Bene}
\affiliation{MTA Atomki Lend\"ulet Quantum Correlations Research Group, Institute for Nuclear Research, P.O. Box 51, H-4001 Debrecen, Hungary} 

\author{Tam\'as V\'ertesi}
\affiliation{MTA Atomki Lend\"ulet Quantum Correlations Research Group, Institute for Nuclear Research, P.O. Box 51, H-4001 Debrecen, Hungary} 

\date{\today}


\begin{abstract}
We address the problem of testing the quantumness of two-dimensional systems in the prepare-and-measure (PM) scenario, using a large number of preparations and a large number of measurement settings, with binary outcome measurements. In this scenario, we introduce constants, which we relate to the Grothendieck constant of order 3. We associate them with the white noise resistance of the prepared qubits and to the critical detection efficiency of the measurements performed. Large-scale numerical tools are used to bound the constants. This allows us to obtain new bounds on the minimum detection efficiency that a setup with 70 preparations and 70 measurement settings can tolerate. 
\end{abstract}

\maketitle

\section{Introduction}
Quantum theory reveals interesting and counter-intuitive phenomena in even the simplest physical systems. Paradigmatic examples are Bell nonlocality~\cite{bell64,brunner_review} and Einstein-Podolsky-Rosen (EPR) steering~\cite{wiseman07,bowles14,sania14,uola}. These nonlocal phenomena appear as strong correlations between the outcomes of spatially separated measurements performed by independent observers. These correlations enable us to distinguish the classical and quantum origins of the experiments. Recently, a similar split between classical and quantum features was found in a setup closely related to quantum communication tasks, the so-called prepare-and-measure (PM) scenario~\cite{Gallego}. This scenario can be viewed as a communication game~\cite{Buhrman} between two parties, Alice (the sender) and Bob (the receiver), where the dimension of the classical (versus quantum) system communicated from Alice to Bob is bounded from above. 

\begin{center}
\begin{figure}[t!]
\includegraphics[trim=-20 10 25 0,clip,width=11.5cm]{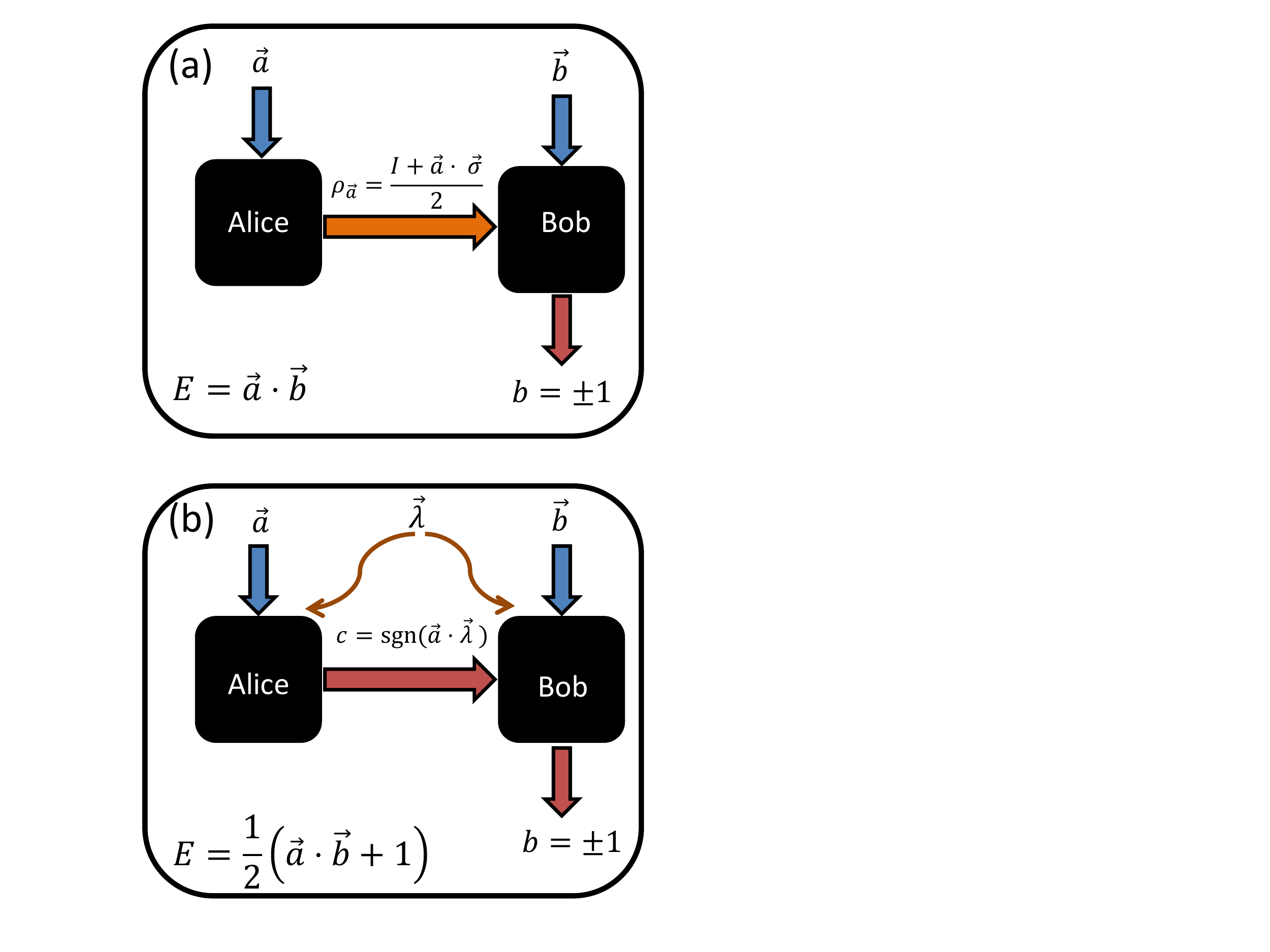}
\caption{The prepare-and-measure setup for (a) qubit communication and (b) a classical model using one bit of communication. In (a) upon receiving the input settings $\vec a$ and $\vec b$, Alice sends to Bob a qubit in the quantum state $\rho_{\vec a}$. Then Bob performs a projective measurement $M_{b|\vec b}=(\one+b\vec b\cdot\sigma)/2$, where the two outcomes are labelled by $b=\pm 1$. As a result, the  expectation value of Bob's $\pm 1$ outcome becomes $E(\vec a,\vec b)=\vec a\cdot\vec b$ (see equation~(\ref{Exydot})).
In (b) the classical one bit Gisin-Gisin protocol~\cite{GG} is as follows. The shared randomness $\vec\lambda$ is distributed between the two parties, where the unit vector $\vec\lambda\in S^2$ is chosen uniformly at random from the sphere. After obtaining the settings $\vec a$ and $\vec b$, Alice communicates to Bob the classical binary message $c=\text{sgn}(\vec a\cdot\vec\lambda)$. Then Bob outputs $b=\text{sgn}(c\vec b\cdot\vec\lambda)$ with probability $|\vec b\cdot\vec\lambda|$, and $b=0$ with probability $1-|\vec b\cdot\vec\lambda|$. Finally, Bob performs a coarse graining on his outputs by grouping $b=0$ with $b=1$ and identifying both of them with outcome $b=1$. As a result, as it is shown in Sec.~\ref{sec:gisinmodel}, the expectation value of Bob's $b=\pm 1$ outcome becomes $E(\vec a,\vec b)=(\vec a\cdot\vec b+1)/2$.} \label{fig:setup}
\end{figure}
\end{center}

The PM game is described as follows (see panel (a) of Fig.~\ref{fig:setup}). Upon receiving an input $x=(1,\ldots,n)$, a preparation device (controlled by Alice) emits a physical system in a quantum state $\rho_x$. We assume $\rho_x\in\mathcal L(\CC^d)$ for a given $d\ge 2$. In the following, however, we will focus explicitly on $d=2$, that is, we assume that two-dimensional quantum systems (qubits) or classical systems (bits) are transmitted from Alice to Bob. The state $\rho_x$ is passed to a measurement device which, upon receiving an input $y=(1,\ldots,m)$ performs a measurement and obtains an outcome $b=(1,\ldots,o)$.  In this paper we will focus on the smallest, nontrivial case of $o=2$, i.e., measurements with two outcomes, in which case we denote the outcomes by $b=\pm 1$. 

Our goal in this scenario is to compare and quantify the performance of qubits with that of classical bits. This scenario has been discussed to some extent for a small number of preparations $n$ and measurements $m$ (see e.g. Refs.~ \cite{Gallego,Ahrens12,Hendrych12,Ahrens14,Poderini20,Gois21,Drotos23}. Note also that the emblematic protocol, the so-called quantum random access code~\cite{QRAC} (QRAC), is a special instance of the PM game. See Ref.~\cite{Buhrman} for more references on communication protocols related to QRAC. These games have also found applications in randomness generation (see~\cite{Li12,Mannalath22}). More recent notable generalizations of QRAC protocols have been considered in Refs.~\cite{Vaisakh,Patra,DM21,Alves23}. 

However, in this paper we would like to turn our attention to the case of large $n$ and $m$ (i.e. in the range of 70). We will see that the main bottleneck of the study is the computation of the relevant quantities associated with the classical bit case for which we develop large scale numerical tools in this paper. We first concentrate on the qubit case, and then we will elaborate on the classical bit case. In the qubit case we define $q(M)$, whereas in the classical bit case we define the quantities $S(M)$ and $L_2(M)$. These quantities in turn define the ratios $q(M)/L_2(M)$ and $(q(M)-S(M))/(L_2(M)-S(M))$, which upper-bound our new constants $K_{\text{PM}}$ and $K_{\text D}$, respectively. These constants have the physical meaning of defining the respective critical white noise tolerance and critical detection efficiency of the binary-outcome measurements in the qubit prepare-and-measure scenario.  

In this paper, we relate these two introduced constants to the purely mathematical Grothendieck constant, $K_G$~\cite{Grot}. More generally, Grothendieck’s problem has implications for many areas of mathematics. It first had a major impact on the theory of Banach spaces and then on $\mathcal{C}^\ast$-algebras. More recently, it has influenced graph theory and computer science (see e.g.~\cite{Khot}). Furthermore, a connection of the Grothendieck problem to Bell nonlocality was noticed by Tsirelson~\cite{tsirelson93}. Subsequently, Acin et al.~\cite{acin06}, based on the work of Tsirelson, exploited this connection to show that the critical visibility of the Bell nonlocal two-qubit Werner state is given by $1/K_G(3)$, where $K_G(3)$ is a refined version of Grothendieck's constant~\cite{krivine}. Relating the local bound of correlation Bell scenarios to the classical bit bound of PM communication scenarios, we find in this paper that the new constant $K_{\text{PM}}$ is equal to $K_G(3)$. We also introduce the constant $K_{\text D}$, which we relate to the critical detection efficiency $\eta_{\text{crit}}$ of binary-outcome measurements in the qubit PM scenario. In particular, we find in our model for finite detection efficiency that $\eta_{\text{crit}}=1/K_{\text D}$. Armed with our efficient numerical tools, we bound the constant $K_{\text D}$ from below, which implies an upper bound of $0.6377$ on $\eta_{\text{crit}}$.

{\it Qubit case.}---In the qubit binary outcome ($o=2$) case, the measurement is described by two positive operators $\{\Pi_{b|y}\}$, $b=\pm 1$ acting on $\CC^2$ which sum to the identity $\Pi_{b=+1|y}+\Pi_{b=-1|y}=\one$ for each $y$, where $\one$ denotes the $2\times 2$ identity matrix. The statistics of the experiment are then given by the formula
\begin{equation}
P(b|x,y)=\Tr(\rho_x \Pi_{b|y}).
\label{born}
\end{equation}
It is important to note that both the preparations and the measurements are unknown to the observer, up to the fact that the dimension of the transmitted system is two. Since we have binary outcomes $b=\{+1,-1\}$ it becomes convenient to use expectation values
\begin{equation}
E_{x,y}=P(b=+1|x,y)-P(b=-1|x,y).
\label{Exy}
\end{equation}
Note that $E_{x,y}$ can take up the values in $\left[-1,+1\right]$ for all $x,y$. However, if the Hilbert space dimension of the communicated particle is bounded, then in general not all expectation values $E_{x,y}$ in $\left[-1,+1\right]$ become possible. The simplest scenario that shows this effect appears already for $n=3$, $m=2$ and $o=2$ (see~\cite{Gallego} for an example).

With respect to the measurement operators $M_{b|y}$, one case, namely the set of projective rank-1 measurements, is of particular interest to us. In this case, we have
\begin{equation}
\Pi_{b|y}=\frac{\one_2+b\vec b_y\cdot\vec\sigma}{2},
\end{equation}
where $\vec b_y\in S^2$, $b=\pm 1$ and $\vec\sigma=(\sigma_x,\sigma_y,\sigma_z)$ is the vector of Hermitian $2\times 2$ Pauli matrices. On the other hand, let us set 
\begin{equation}
\rho_x = \frac{\one+\vec a_x\cdot\vec\sigma}{2},
\end{equation}
where $\vec a_x\in S^2$. This density matrix corresponds to a pure state with Bloch vector $\vec a_x$. Note that in this case, the above equations give us
\begin{equation}
\label{Exydot}
E_{x,y}=\vec a_x\cdot\vec b_y,
\end{equation}
where $\vec a_x,\vec b_y\in S^2$.

Limits on the set of possible distributions in dimension two can be captured by the following expression
\begin{equation}
W=\sum_{x=1}^{n}\sum_{y=1}^{m} M_{x,y}E_{x,y},
\label{dimwit}
\end{equation}
where $M_{x,y}$ are coefficients of a real witness matrix $M$ of dimension $n\times m$. Let us then define the quantity
\begin{equation}
Q(M)=\max\sum_{x=1}^{n}\sum_{y=1}^{m}M_{x,y}E_{x,y},
\label{QM}
\end{equation}
where $E_{x,y}$ is of the form (\ref{Exy}), and where we maximize the expression over Bob's measurements $\{M_{b|y}\}$ and the qubit state $\rho_x$ in Eq.~(\ref{born}). Thus, $Q(M)$ is the value that is achievable with the most general two-dimensional quantum resources in our PM setup.
We further define the quantity
\begin{equation}
q(M)=\max\sum_{x=1}^{n}\sum_{y=1}^{m}M_{x,y}E_{x,y},
\label{qM}
\end{equation}
where $E_{x,y}=\vec a_x\cdot\vec b_y$ and we maximize over the unit vectors $\vec a_x$ and $\vec b_y$ in the three-dimensional Euclidean space. It turns out that $Q(M)$ can be obtained with pure qubit states and projective measurements~\cite{Ahrens14}. However, the optimal projective measurements are in general not of rank-1, they can be of rank-0 or rank-2 as well. Indeed, there are example matrices $M$ (even in the simple $n=m=3$, $o=2$ case) for which $Q(M)>q(M)$. Note that $q(M)$ corresponds to projective qubit measurements of rank 1, in which case $E_{x,y}=\vec a_x\cdot\vec b_y$ (see equation~(\ref{Exydot})). Yet, as we will see, the set $\{E_{x,y}\}_{x,y}$ obtained by rank-1 projective measurements is a significant subset of the set $\{E_{x,y}\}_{x,y}$ corresponding to the most general qubit measurements. The tools for computing the value $Q(M)$ can be found in Refs.~\cite{NV15,TRR19}.

Importantly, the value of $Q(M)$ can serve as a dimension witness in the prepare-and-measure scenario~\cite{Gallego}. Indeed, if $W>Q(M)$ for some $M$ (where the witness $W$ is defined by equation~(\ref{dimwit})), this implies that the set of states $\{\rho_x\}_{x=1}^{n}$ transmitted to Bob must have contained at least one state $\rho_{x=x'}$ of at least three dimensions (that is qutrit). 

{\it Classical bit versus qubit case.}---It turns out that the witness $W$ can also serve as a quantumness witness. To this end, let us discuss the classical bit case. That is, we want to bound the expression~(\ref{dimwit}) if Alice can only prepare classical two-dimensional systems (i.e. bits). Let us denote the bound on (\ref{dimwit}) by $L_2(M)$, which corresponds to this situation. If $W>L_2(M)$, this certifies that some of the measurements performed by Bob are true (incompatible) quantum measurements acting on true qubit states~\cite{Gallego,QRAC_selftest}. 
Mathematically, the classical bit case is equivalent to the qubit case discussed above, with the restriction that all qubits are sent in the same basis, and all measurements of Bob are carried out in the very same basis. That is, if we want to maximize (\ref{dimwit}) for correlations $E_{x,y}$ arising from classical two-dimensional systems, the maximum can be attained with pure states
\begin{equation}
\rho_x = \frac{\one_2+a_x\cdot\sigma_z}{2},
\label{rhoxbit}
\end{equation}
where $a_x=\pm 1$, and observables $B_y=\Pi_{0|y}-\Pi_{1|y}$ which have the form 
\begin{equation}
B_y=b_y^+\ket{0}\bra{0}+b_y^-\ket{1}\bra{1},
\label{Bybit}
\end{equation}
where $\sigma_z$ is the standard Pauli matrix 
\begin{align*}
\sigma_z=\begin{pmatrix*}[r]
	1 & 0 \\
	0& -1
\end{pmatrix*}
\end{align*}
and both $b_y^+$, $b_y^-$ are $\pm 1$ variables. Inserting these values into (\ref{Exy}) we obtain
\begin{equation}
E_{x,y} = \frac{(1+a_x)b_y^++(1-a_x)b_y^-}{2}.
\label{Exybit}
\end{equation} 
Since we have binary variables $a_x=\pm 1$, they translate to $E_{x,y}=b_y^+$ if $a_x=1$ and $E_{x,y}=b_y^-$ if $a_x=-1$. Then the classical one-bit bound $L_2(M)$ is given by
\begin{equation}
L_2(M)=\max\sum_{x=1}^{n}\sum_{y=1}^{m}M_{x,y}E_{x,y},
\label{L2M}
\end{equation}
where $E_{x,y}$ is defined by (\ref{Exybit}) and we maximize over all binary variables $a_x, b^+_y,b^-_y\in\{-1,+1\}$. In words, the expression~(\ref{Exybit}) corresponds to the following deterministic protocol. Alice, depending on $x$, prepares a bit $a_x=\pm 1$, which she sends to Bob, who outputs $b=\pm 1$ depending on the value of $a_x$ and the measurement setting $y$. That is, Bob's output is a deterministic function $b=f(a_x,y)$, where the output assumes $b=\pm 1$. We can write
\begin{align} 
&L_2(M)\nonumber\\
&=\max \left(\sum_{x: a_x=+1}\sum_{y=1}^{m} M_{xy}b^+_y+\sum_{x: a_x=-1}\sum_{y=1}^{m} M_{xy}b^-_y\right),
\label{L2Mv2}
\end{align}
where the maximum is taken over all binary $a_x$, $b_y^+$ and $b_y^-$ variables $\pm 1$. We can eliminate the variables $b^+_y$ and $b^-_y$ from the above expression and get the following formula for $L_2(M)$:
\begin{equation}
L_2(M) = \max_{a_x=\pm 1} \left(\left\|\sum_{x: a_x=+1} M_x\right\|_1+\left\|\sum_{x: a_x=-1} M_x\right\|_1\right),
\label{L2Mv3}
\end{equation}
which only consists of maximization over the binary variables $a_x=\pm 1$. In the above formula, $M_x$ denotes the $x$th row of the real $n\times m$ matrix $M$, where \(\|v\|_1\) denotes the Manhattan norm of the real vector $v$, i.e., $\|v\|_1=\sum_x |v_x|$. We prove several interesting properties of $L_2(M)$ in the Methods section~\ref{sec:L2proofs}. In particular, $L_2$  is proven to be a matrix norm. Let us recall that $L_2(M)$ is a key quantity in our study, as it enables witnessing both quantumness of preparations and quantumness of measurements. Indeed, $W>L_2(M)$, where $W$ is defined in equation~(\ref{dimwit}), certifies incompatible quantum measurements acting on true qubit states. That is, not all the performed measurements and not all prepared states originate from the same basis~\cite{Gallego}. In section~\ref{sec:L2proofs} we do not restrict our study to the properties of the $L_2$ norm but generalize $L_2(M)$ to $L_k(M)$ for any $k>2$ and prove that $L_k$ is a norm as well, moreover $L_k(M)$ is a monotonic increasing function of $k$. Furthermore, in section~\ref{sec:L2tips} we give tips for an efficient implementation of the branch-and-bound algorithm~\cite{BB} for computing the $L_k(M)$ bound for $k=2$ and for $k>2$ as well. 

{\it Introducing the constants $K_{\text{PM}}$ and $K_{\text{D}}$.}---We define two quantities $K_{\text{PM}}$ and $K_{\text{D}}$ which are related to $L_2(M)$ and $q(M)$, and are defined as follows. Let us first introduce $K_{\text{PM}}$, in which case we ask for the maximum ratio between $q(M)$ and $L_2(M)$. That is, we are interested in the value
\begin{equation}
K_{\text{PM}}=\max_M \frac{q(M)}{L_2(M)},
\label{KPM}
\end{equation}
where the maximization is taken over all possible real $n\times m$ matrices $M$, where $q(M)$ is defined by (\ref{qM}) and $L_2(M)$ is defined by~(\ref{L2M}). 

Let us now recall the Grothendieck constant of order 3~\cite{Grot,krivine,acin06,pisier12,hua15}, which is given by 
\begin{equation}
K_G(3)=\max_M \frac{q(M)}{L(M)},
\label{KG3}
\end{equation}
where the maximization is taken over real matrices $M$ of arbitrary dimensions $n\times m$, $q(M)$ is defined by (\ref{qM}) and $L(M)$ is defined as follows
\begin{equation} 
L(M)=\max \sum_{x=1}^{n}\sum_{y=1}^{m} M_{x,y}a_xb_y,
\label{LMdef}
\end{equation}
where the maximum is taken over all $a_x, b_y\in\{-1,+1\}$. The value of $K_G(3)$ in (\ref{KG3}), according to the recent work of Designolle et al.~\cite{designolle},  is bounded by
\begin{equation}
\label{KG3bounds}
1.4367\le K_G(3)\le 1.4546,
\end{equation}
where the lower bound is an improved version of that given in Ref.~\cite{grot4} and the upper bound is an improved version of that given in Refs.~\cite{finch,KG3}. See Ref.~\cite{wiki} for some historical data on the best lower and upper bounds for $K_G(d)$. We prove that $K_{\text{PM}}=K_G(3)$, which will be given in the Results section~\ref{sec:KPMeqKG3}. We are interested in $K_{\text{D}}$ as well, a quantity similar to $K_{\text{PM}}$. We define this quantity as follows
\begin{equation}
K_{\text{D}}=\max_M \frac{q(M)-S(M)}{L_2(M)-S(M)},
\label{KD}
\end{equation}
where 
\begin{equation}
S(M)=\sum_{x=1}^n\sum_{y=1}^m M_{x,y}.
\label{SM}
\end{equation}
Note the relation
\begin{equation}
\frac{q(M)-S(M)}{L_2(M)-S(M)}\ge \frac{q(M)}{L_2(M)},
\end{equation}
whenever $L_2(M)>S(M)$ (also note that $q(M)\ge L_2(M)$), therefore we have $K_{\text{D}}\ge K_{\text{PM}}=K_G(3)$. From this we immediately obtain the lower bound $K_{\text{D}}\ge 1.4367$. In this paper, we give efficient large-scale numerical methods to obtain even better lower bounds on the above quantity. Namely, we prove the lower bound $K_{\text{D}}\ge 1.5682$. We also prove an upper bound of 2 on this quantity, so putting all together we have the following interval
\begin{equation}
1.5682\le K_{\text{D}}\le 2
\label{KD_LU}
\end{equation}
for the constant $K_{\text{D}}$. It is an open problem to close or at least reduce the gap between the lower and upper limits.

We next present the Results section, which contains our main findings in three subsections.

\section{Results}
\label{sec:results}

\subsection{Proof of the relation \texorpdfstring{$K_{\text{PM}}=K_G(3)$}{KPMKG3}}
\label{sec:KPMeqKG3}

To prove our claim, we relate $L(M')$ to $L_2(M')$, where $M'$ is given by the following matrix (see also~(\ref{Mv}))
\begin{align}
M'=\begin{pmatrix*}
	&M \\
 -&M 
\end{pmatrix*},
\label{Mvv}
\end{align}
where $M$ is a real $n\times m$ matrix. Denote by $M_x$ the $x$-th row of the matrix $M$. Note that according to the above definition $M'$ has size $2n\times m$ and $M'$ has rows such that $M'_x=M_x$ and $M'_{x+n}=-M_{x}$ for all $x=1,\ldots,n$. Then the following lemma holds.

\begin{lemma}
\label{L2L}
$L_2(M')=L(M')=2L(M)$ for any matrix $M'$ of the form (\ref{Mvv}), where $L_2$ is the $L_2$ norm given by the definition (\ref{L2M}) and $L$ is the local bound given by (\ref{LMdef},\ref{LDef}).
\end{lemma}
The proof of this lemma is given in Methods section~\ref{sec:LL2Mp}.  Then we need to prove the following lemma.

\begin{lemma} 
$K_{\text{PM}}\le K_G(3)$.
\label{lemma_eq1}
\end{lemma}
For an arbitrary matrix $M$, we have $L_2(M)\ge L(M)$. This has been proved in Methods section~\ref{sec:L2proofs}. Then the lemma follows from the definitions~(\ref{KG3}) and (\ref{KPM}). Our next lemma reads
\begin{lemma} 
$K_{\text{PM}}\ge K_G(3)$.
\label{lemma_eq2}
\end{lemma}

{\noindent\it Proof.}
To prove this, it suffices to show that for an arbitrary real matrix $M$, there exists the matrix $M'$ defined by (\ref{Mvv}) such that $q(M')=2q(M)$ and $L_2(M')=2L(M)$. The first relation follows from the special structure of $M'$. The second relation has been shown in Lemma~\ref{L2L}. Therefore, $K_{\text{PM}}$ cannot be less than $K_G(3)$, which proves our claim. $\qed$

\begin{corollary}
As a corollary of the above Lemmas~\ref{lemma_eq1} and~\ref{lemma_eq2} we obtain $K_{\text{PM}}=K_G(3)$.
\end{corollary}

Hence we have the same bounds $1.4367\le K_{\text{PM}}\le 1.4546$ as for $K_G(3)$ (see~(\ref{KG3bounds})). From the corollary above, we have a matrix $M'$ of size $48\times 24$ with $q(M')/L_2(M')>\sqrt 2$. Indeed, the construction is based on a matrix $M$ of size $24\times 24$, which provides $q(M)/L(M)>\sqrt 2$~\cite{Gilbert}. To the best of our knowledge, this is the smallest $M$ matrix that has the property $q(M)/L(M)>\sqrt 2$. Then the $48$-by-$24$ matrix $M'$ follows from~(\ref{Mvv}). On the other hand, $q(M)/L(M)=\sqrt 2$ is already attained with a $2\times 2$ matrix $M$ in the CHSH-form~\cite{chsh69}:
\begin{align*}
M=\begin{pmatrix*}[r]
	1 & 1 \\
	1& -1
\end{pmatrix*}.
\end{align*}
It remains an open question to show that $K_{\text{PM}}>\sqrt 2$ with a matrix size smaller than $48\times 24$, which might use a different construction than the one above.

\subsection{Proof of the bounds \texorpdfstring{$1.5682\le K_{\text{D}}\le 2$}{KD}}
\label{sec:uplow}

{\it Upper bound.}---We first prove the upper bound. Translating the Gisin-Gisin model~\cite{GG} from the Bell nonlocality~\cite{bell64,brunner_review} to the PM scenario~\cite{Gallego}, we find that the following statistics can be obtained in the PM scenario with 1 bit of classical communication: 
\begin{equation}
\label{Egg}
E(\vec a,\vec b)=\frac{\vec a\cdot\vec b + 1}{2},
\end{equation}
where $\vec a\in S^2$ denotes the preparation vector and $\vec b\in S^2$ denotes the measurement Bloch vector. We give the proof of this formula in the Methods section~\ref{sec:gisinmodel} and we show panel (b) in figure~\ref{fig:setup} for the description of the classical one-bit model. 
On one hand, due to the above Gisin-Gisin one-bit model, we have for an arbitrary $n\times m$ matrix $M$:
\begin{equation}
\max\sum_{x=1}^{n}\sum_{y=1}^{m}M_{x,y}E_{x,y}\le L_2(M),
\label{giga1}
\end{equation}
where $E_{x,y}$ has the form~(\ref{Egg}) and we maximized over the unit vectors $\vec a_x$ and $\vec b_y$ in the three-dimensional Euclidean space. On the other hand, substituting $E_{x,y}:=E(\vec a_x,\vec b_y)$ in the formula~(\ref{Egg}) into (\ref{giga1}) we find
\begin{align}
\max\sum_{x=1}^{n}\sum_{y=1}^{m}M_{x,y}E_{x,y} &= \frac{\max\sum_{xy} M_{xy}\left(\vec a_x\cdot\vec b_y + 1\right)}{2}\nonumber\\
&= \frac{q(M) +S(M)}{2},
\label{giga2}
\end{align}
where maximization is over the unit vectors $\vec a_x$ and $\vec b_y$ in the three-dimensional Euclidean space, and we also used the definition of $q(M)$ in (\ref{qM}) and the definition of $S(M)$ in (\ref{SM}). Comparing the right-hand side of (\ref{giga1}) with (\ref{giga2}), we have 
\begin{equation}
\frac{q(M)-S(M)}{L_2(M)-S(M)}\le 2,
\end{equation}
where the left-hand side of (\ref{KD}) is just $K_{\text{D}}$, which proves the upper bound $K_{\text{D}}\le 2$. \qed

{\it Lower bound.}---In the following, we prove the lower bound using large-scale numerical tools. Note, however, that the resulting bound is rigorous and in particular the final result is due to exact computations. The steps are as follows.

Given a fixed setup with Alice's Bloch vectors $\vec a_x$, $x=(1,\ldots,n)$ and Bob's Bloch vectors $\vec b_y$, $y=(1,\ldots,m)$ the method is the following. We define the $(n\times m)$-dimensional one-parameter family of matrices $E_{xy}(\eta)$ with entries
\begin{equation}
E_{x,y}(\eta)=\eta E_{x,y}+(1-\eta),
\label{Exyeta}
\end{equation}
where $E_{x,y}=\vec a_x\cdot\vec b_y$. We wish to show that for some $\eta\in\left[0,1\right]$, the distribution~(\ref{Exyeta}) in the PM scenario cannot be simulated with one bit of classical communication. In fact, due to the expectation value~(\ref{Egg}) of the Gisin-Gisin model, it is enough to consider the interval $\eta\in\left[1/2,1\right]$. To show quantumness, we therefore need to find a matrix $M$ of certain size $n\times m$ and a given $\eta\in\left[1/2,1\right]$ such that 
\begin{equation}
\label{hyper}
\sum_{x=1}^n\sum_{y=1}^m M_{xy}E_{xy}(\eta)>L_2(M)
\end{equation}
for $E_{xy}(\eta)$ defined by (\ref{Exyeta}), and $L_2(M)$ is defined by (\ref{L2M}).  The above problem, i.e., finding a suitable $M$ with the smallest possible $\eta$ in (\ref{hyper}), can be solved by a modified version~\cite{Gilbert} of the original Gilbert algorithm~\cite{Gilb66}, a popular collision detection method used, for example, in the video game industry.

The algorithm is iterative, and the procedure adapted to our problem is given in Sec.~\ref{sec:gibi}. Indeed, using the algorithm of Gilbert, we find the value 
\begin{equation}
\eta^*=0.6377
\end{equation}
and a corresponding $70\times 70$ matrix $M$ and $E_{xy}(\eta^*)$ in the form~(\ref{Exyeta}) which satisfies inequality~(\ref{hyper}). We will give more technical details of the input parameters and the implementation of the algorithm in Section~\ref{sec:techgibi}.  
Then, rearranging (\ref{hyper}) and making use of equation~(\ref{Exyeta}), we find the bound
\begin{equation}
\frac{\sum_{xy}M_{xy}E_{xy}-S(M)}{L_2(M)-S(M)}>\frac{1}{\eta^*},
\end{equation}
where due to the definitions~(\ref{qM},\ref{KD}) 
the lower bound
\begin{equation}
K_{\text{D}}>(1/\eta^*)\simeq1.5682
\end{equation}
 on $K_{\text{D}}$ follows.

\subsection{Physical meaning of the constants \texorpdfstring{$K_{\text{PM}}$}{KPM} and \texorpdfstring{$K_{\text{D}}$}{KD}}

{\it The role of $K_{\text{PM}}$ in the PM scenario.}---The value of $K_{\text{PM}}$ is interesting from a physical point of view as well, since it is related to the critical noise resistance of the experimental setup if the transmitted $\rho_x$ goes through a noisy, fully depolarizing channel. That is, $1-p_{\text{crit}}=1-(1/K_{\text{PM}})$ gives the amount $(1-p_{\text{crit}})$ of critical white noise $\one/2$ that the PM experiment with rank-1 projective qubit measurements can maximally tolerate while still being able to detect quantumness. Namely, for a fully depolarizing channel with visibility parameter $p$ the qubits $\rho_x$ emitted by Alice turn into $p\rho_x+(1-p)\one/2$, and the expectation value~(\ref{Exydot}) becomes
\begin{equation}
E_{xy}=p\vec a_x\cdot\vec b_y,
\label{Exyp}
\end{equation}
where $\{\vec a_x\}_x$ are the Bloch vectors of Alice's qubits, whereas $\{\vec b_y\}_y$ are the Bloch vectors of Bob's measurements. To witness quantumness, there must exist expectation values $E_{xy}$ in (\ref{Exyp}) and a matrix $M$ of arbitrary size such that 
\begin{equation} 
\sum_{xy}M_{xy}E_{xy}>L_2(M).
\label{Witp}
\end{equation}
Inserting (\ref{Exyp}) into (\ref{Witp}) and making use of (\ref{qM}), we obtain
\begin{equation}
\label{pcrit}
p_{\text{crit}}=\min_M\frac{L_2(M)}{q(M)}=\frac{1}{K_{\text{PM}}}=\frac{1}{K_G(3)}
\end{equation}
for the critical noise tolerance. In fact, the value of $K_G(3)$ appears in the studies~\cite{acin06,vertesi_gro,KG3} of the Bell nonlocality of two-qubit Werner states~\cite{werner89}. 
Note that a recent approach in Ref.~\cite{bowles15}, based on the simulability of Werner states with local models, yields the same relation~(\ref{pcrit}) between $p_{\text{crit}}$ and $1/K_G(3)$ . 

From the upper and lower bounds on $K_{\text{PM}}$, the following bounds on the amount $(1-p_{\text{crit}})$ of critical white noise follow:
\begin{equation}
0.3039\le (1-p_{\text{crit}})\le 0.3125.
\end{equation}

{\it The role of $K_{\text{D}}$ in the PM scenario.}---In section~\ref{sec:uplow} we proved the lower bound of $K_{\text{D}}\ge  1.5682$. Below we prove that this bound is related to the finite detection efficiency threshold of Bob's measurements. 
To this end, we assume that Bob's detectors are not perfect and only fire with probability $\eta$. Assume that when the measurement $y$ fails to detect, Bob outputs $b_y=1$ (due to possible relabelings there is no loss of generality). Assume further that the probability of detection $\eta$ is the same for all $y$. This is the problem of symmetric detection efficiency. A review of this problem in the Bell scenario can be found in Ref.~\cite{Larsson}. On the other hand, the same problem in the PM scenario has been elaborated in Refs.~\cite{arno12,li15} and the upper bound of $1/\sqrt 2$ on the critical value of the symmetric detection efficiency was found. 

Since $\eta$ does not depend on $y$, the expectation value $E_{x,y}$ becomes $E_{x,y}(\eta)=\eta E_{x,y} + (1-\eta)$ for all $x$ and $y$. Hence, the witness matrix $M$ detects quantumness with finite detection efficiency $\eta$ (assuming optimal preparation states and measurements) whenever we have
\begin{equation}
\eta q(M) + (1-\eta)\sum_{x,y}M_{x,y}>L_2(M).
\label{etaineq}
\end{equation}
Recalling $S(M)=\sum_{x,y}M_{x,y}$, solving the above relation for $\eta$, and optimizing over all $M$ witness matrices, we obtain the critical detection efficiency $\eta_{\text{crit}}$:
\begin{equation}
\eta_{\text{crit}}=\min_M{\frac{L_2(M)-S(M)}{q(M)-S(M)}}=\frac{1}{K_{\text{D}}},
\label{etacrit}
\end{equation}
where $K_{\text{D}}$ is defined by (\ref{KD}). In particular using the lower bound value $K_{\text{D}}\ge 1.5682$, we obtain the improved upper bound $0.6377$ on $\eta_{\text{crit}}$.

It should be noted, however, that the above is not the most general detection efficiency model. Rather than outputting $b_y=1$, Bob can output a third result, which could potentially give a lower detection efficiency threshold. An open problem is whether this third outcome can lower the detection efficiency threshold. In the above, we also assumed that Bob's qubit measurements are rank-1 projectors that can achieve $q(M)$. However, it is known that the true qubit maximum $Q(M)$ (in~(\ref{QM})) can be larger than $q(M)$ (in~(\ref{qM})) for a given $M$. Hence, we can say that the most general symmetric detection efficiency threshold is upper bounded by $1/K_{\text{D}}$, and it is an open problem whether this upper bound is tight or not.

Let us mention that in the two-outcome scenario a different type of modelling of the loss event due to the finite detection efficiency can also be imagined. Namely, let us assume that Bob associates the outcomes $+1$ and $-1$ to the no-detection event with equal probability. In this case, the expectation value $E_{x,y}(\eta)=\eta E_{x,y}+(1-\eta)$ when outcome $+1$ is assigned to the no-detection event becomes $\eta E_{x,y}$. This leads to the modified inequality $\eta q(M)>L_2(M)$ in Eq.~(\ref{etaineq}) and the modified critical detection efficiency, $\eta_{\text{crit}}=\min_M{(L_2(M)/q(M))}=1/K_{\text{PM}}$. Therefore, using Bob's non-deterministic assignment of the $\pm 1$ outcomes for the no-detection event, the critical detection efficiency can be linked to $K_G(3)=K_{\text{PM}}$, i.e., the Grothendieck constant of order 3. Note, however, that due to our finding that $K_G(3)<K_{\text{D}}$, the critical detection efficiency in this non-deterministic modelling of the no-detection event will be suboptimal compared to the deterministic assignment model, when we associate the no-detection event with a given outcome.  
\section{Methods}
\label{Methodssec}
\subsection{Properties of the \texorpdfstring{$L_2$}{L2} and \texorpdfstring{$L_k$, $k>2$}{Lk} norm}
\label{sec:L2proofs}

{\it Notations.}---We first introduce notation used throughout this subsection. Let \(A^n, n = 0, 1, 2, \dots\) be the set of \(n\) dimensional vectors over the set \(A\). Let $v_i$ denote the $i$th element of $v \in A^n$ ($i = 1, 2, \dots, n$). Let \(\_;\_ : A^n \times A^m \rightarrow A^{n+m}\) denote the
concatenation of vectors. Let \(()\) the singleton element of \(A^0\). Further let \((a) \in A^1\) if \(a \in A\). The parenthesis may be omitted so
\((1); (2); (3) = 1; 2; 3 \in \mathbb R^3\), for example. Let \(\overline a^{n} = a; a; ...; a \in A^n\) where \(a \in A\). We
write \(\overline a\) instead of \(\overline a^{n}\) if \(n\) can be inferred from the context. We define \(\Mat{n}{m}\) as the set of real \(n\times m\) matrices.
Matrices are represented as vectors of their row vectors, i.e.~\(\Mat{n}{m} = (\mathbb R^m)^n\). Let $\transp{M} \in \Mat{m}{n}$ be the transposition of $M \in \Mat{n}{m}$ and let $\Id{m} \in \Mat{m}{m}$ denote the $m\times m$ identity matrix. Further, it is convenient to define by $\W{n}{k} = \{\Id{k}_j \,|\, j = 1, 2, \dots, k\}^n \subset \Mat{n}{k}$ the set of matrices whose rows are all $0$s, but exactly one is $1$. Let $\Perm{n} \subset\Mat{n}{n}$ denote the set of permutation matrices. Let $\SM{M} = \sum_{i=1}^n \left\|M_i\right\|_1$ denote the Manhattan norm of the matrix $M \in \Mat{n}{m}$.

{\it Definition of $L_k$.}---We first give the definition of $L_k$. Let $k \in \mathbb N^{+}.$
\begin{equation}
\begin{aligned}
&L_k : \Mat{n}{m} \rightarrow \mathbb R\\
&L_k(M) = \max_{W \in \W{n}{k}} \SM{\transp{W}M}.
\end{aligned}
\label{L2Def}
\end{equation}
Note that $W$ is defined above in {\it Notations} and $\transp{W}$ denotes the transposed matrix of $W$.  
We prove below that Eq.~(\ref{L2Def}) corresponds to Eq.~(\ref{L2Mv3}) in the case of $k=2$. The proof is as follows
\begin{align*}
\max_{a_x\in\{\pm 1\}}& \left(\Man{\sum_{x: a_x=+1} M_x} + \Man{\sum_{x: a_x=-1} M_x}\right)\\
&= \max_{W\in\W{n}{2}} \left(\Man{\transp{W}_1M}+ \Man{\transp{W}_2M}\right)\\
&= \max_{W\in\W{n}{2}} \SM{\transp{W}M}\\
&= L_2(M).
\end{align*}

{\it Properties of $L_k$.}---We prove several interesting properties of $L_k$. Note that our focus in the main text is on $k=2$.  However, the general case $k\ge 2$ is of interest for its own sake. Moreover, it is also motivated physically, corresponding to classical communication beyond bits~\cite{Gallego,brunner13}. First we prove that $L_k$ is a norm for any $k\ge 2$. To this end, we prove its homogeneity, positive definiteness and subadditivity properties.

\begin{lemma}
\(L_k\) is a norm.
\end{lemma}

\noindent Homogeneity:
\begin{align*}
L_k(tM)&= \max_{W \in \W{n}{k}} \SM{\transp{W}(tM)}\\
&= \max_{W \in \W{n}{k}} |t|\SM{\transp{W}M}\\
&= |t|\max_{W \in \W{n}{k}} \SM{\transp{W}M}\\
&= |t|L_k(M),
\end{align*}
\noindent where $|t|$ denotes the absolute value of the scalar $t$ and $L_k$ is defined by (\ref{L2Def}).

\noindent Positive definiteness:
\begin{align*}
L_k(M) &= 0\\
&\Rightarrow\max_{W \in \W{n}{k}} \SM{\transp{W}M} = 0\\
&\Rightarrow \forall W \in \W{n}{k}: \SM{\transp{W}M} = 0\\
&\Rightarrow \forall W \in \W{n}{k}: \transp{W}M = \overline{\overline 0}\\
&\Rightarrow \forall i: M_i = \overline 0\\
&\Rightarrow M = \overline{\overline 0}.
\end{align*}

\noindent Triangle inequality:
\begin{align*}
&L_k(M+N) = \max_{W \in \W{n}{k}} \SM{\transp{W}(M+N)}\\
&\le \max_{W \in \W{n}{k}} (\SM{\transp{W}M} + \SM{\transp{W}N})\\
&\le \max_{W \in \W{n}{k}} \SM{\transp{W}M} + \max_{W \in \W{n}{k}} \SM{\transp{W}N}\\
&= L_k(M) + L_k(N).
\end{align*}
\qed

Let us define $L(M)$ as follows 
\begin{equation}
L(M) = \max_{v \in \{-1, +1\}^n} \Man{vM}.
\label{LDef}
\end{equation}
The above definition is consistent with the one given in~(\ref{LMdef}). $L(M)$ is the local or classical bound of correlation Bell inequalities~\cite{tsirelson93} defined by the correlation matrix $M$ in~(\ref{LDef}). The $L(M)$ quantity also appears in computer science literature under the name of $K_{m,n}$-quadratic programming~\cite{RS09}. Let us note that recently an efficient computation of $L(M)$ has been proposed in Ref.~\cite{grot4} along with the code~\cite{Lcode}.

First we prove the basic property that $L_2(M)\ge L(M)$ for any $M$. Next we prove that $L_k(M)\le L_{k+1}(M)$ for $k\ge 2$. Then we bound $L_k(M)$ from above by the value of $L(M)$ multiplied by $k$. However, we do not know whether the bound can be saturated or not. The lemma stating our first claim is as follows
\begin{lemma}
\begin{equation}
L(M) \le L_2(M),
\end{equation}
\end{lemma}
\noindent where the proof is given as the following chain of equations plus a single inequality invoked in the fourth line
\begin{align*}
L(M)&= \max_{v \in \{-1, +1\}^n} \Man{vM}\\
&= \max_{v \in \{-1, +1\}^n} \Man{\frac12(\overline 1 + v)M - \frac12(\overline 1 - v)M}\\
&= \max_{W \in \W{n}{2}} \Man{\transp{W}_1M - \transp{W}_2M}\\
&\le \max_{W \in \W{n}{2}} (\Man{\transp{W}_1M} + \Man{\transp{W}_2M})\\
&= \max_{W \in \W{n}{2}} \SM{\transp{W}M}\\
&= L_2(M).
\end{align*}
\qed

\noindent Our next lemma proves that $L_k(M)$ is monotone increasing with $k$. 
\begin{lemma}
\begin{equation}
L_k(M) \le L_{k+1}(M).
\end{equation}
\end{lemma}
The proof is given below as the following chain:
\begin{align*}
L_k(M)&= \max_{W \in \W{n}{k}} \SM{\transp{W}M}\\
&= \max_{W \in \W{n}{k}} \sum_{i=1}^k \Man{(\transp{W}M)_i}\\
&= \max_{W \in \W{n}{k}} (\sum_{i=1}^k \Man{(\transp{W}M)_i} + \Man{\overline 0M})\\
&= \max_{W \in \W{n}{k}} \sum_{i=1}^{k+1} \Man{((\transp{W}; \overline 0)M)_i}\\
&= \max_{W \in \W{n}{k}} \SM{(\transp{W}; \overline 0)M}\\
&\le \max_{W \in \mathcal W_{k+1,n}} \SM{\transp{W}M}\\
&= L_{k+1}(M).
\end{align*}
\qed

\noindent Finally, we prove an upper bound on $L_k(M)$. Our lemma reads as follows
\begin{lemma}
\begin{equation}
L_k(M) \le kL(M)
\end{equation}
\label{LkkL}
\end{lemma}

{\noindent\it Proof.}
\begin{align*}
L_k(M)&=\max_{W \in \W{n}{k}} \SM{\transp{W}M}\\
&= \max_{W \in \W{n}{k}} \sum_{i=1}^k \Man{(\transp{W}M)_i}\\
&= \max_{W \in \W{n}{k}} \sum_{i=1}^k \Man{\transp{W}_iM}\\
&= \max_{W \in \W{n}{k}} \sum_{i=1}^k \frac12\Man{(2\transp{W}_i - \overline 1)M + \overline 1M}\\
&\le \max_{W \in \W{n}{k}} \sum_{i=1}^k \frac12(\Man{(2\transp{W}_i - \overline 1)M} + \Man{\overline 1M})\\
&\le \max_{W \in \W{n}{k}} \sum_{i=1}^k \frac12(L(M) + L(M))\\
&= \max_{W \in \W{n}{k}} \sum_{i=1}^k L(M)\\
&= \max_{W \in \W{n}{k}} kL(M)\\
&= kL(M).
\end{align*}
To arrive at the sixth line, we invoked the definition~(\ref{LDef}).
\qed

It is an open question whether Lemma~\ref{LkkL} is tight or not. However, we can find a family of matrices $M^{(k)}$, $k\ge 2$ such that the ratio $L_k(M^{(k)})/L(M^{(k)})$ tends to infinity with increasing $k$. More formally we have
\begin{lemma}
For all $\varepsilon > 0$ there exists a matrix $M$ and $k>1$ such that
\begin{equation}
\frac{L(M)}{L_k(M)} < \varepsilon.
\end{equation}
\end{lemma}
\noindent The proof is based on an explicit construction of matrices $M^k$, $k=(2,\ldots,\infty)$ defined in Ref.~\cite{VP08}. See also Refs.~\cite{EKB1,EKB2}.

{\noindent\it Proof.}
Let $M^k\in\Mat{k}{2^{k-1}}$, $k = (1, 2, \dots,\infty)$ be a family of matrices such that~\cite{VP08}
\begin{equation}
M^k_{i,j} = (-1)^{\left\lfloor\frac{j}{2^{k-i-1}}\right\rfloor}.
\end{equation}
For example,
\begin{equation}
M^4 = \left(\begin{array}{rrrrrrrr}1&1&1&1&1&1&1&1\\1&1&1&1&-1&-1&-1&-1\\1&1&-1&-1&1&1&-1&-1\\1&-1&1&-1&1&-1&1&-1\end{array}\right).
\end{equation}
Now by explicit calculations we find
\begin{equation}
\frac{L(M^k)}{L_k(M^k)} = \frac{k\left(\begin{array}{c}k-1\\\left\lfloor\frac{k-1}2\right\rfloor\end{array}\right)}{k2^{k-1}} \sim \sqrt\frac2{\pi k}.
\end{equation}
\qed

Note that in the particular case of $k=2$ the matrix $M^{(k)}$ is the CHSH expression~\cite{chsh69}, in which case $L(M^{(2)})=2$ and $L_2(M^{(2)})=4$. Hence, for $k=2$ the upper bound in Lemma~\ref{LkkL} is tight. We conjecture that the bound is not tight for greater values of $k$. 

Finally, we show how $L_2$ and in general $L_k$ behaves with the concatenation $(A; B)$ of two matrices $A$ and $B$, where we defined
\begin{equation}
(A;B) = \left[\begin{array}{r}A\\B\end{array}\right].
\end{equation}

\begin{lemma}
Let $A\in\Mat{i}{m}, \; B\in\Mat{j}{m}$. Then we have
\begin{equation}
L_k(A; B) \le L_k(A) + L_k(B).
\end{equation}
\end{lemma}

{\noindent\it Proof.}
\begin{align*}
L_k(A; B)&= \max_{W \in \W{i+j}{k}} \SM{\transp{W}(A; B)}\\
&= \max_{S \in \W{i}{k}} \max_{T\in\W{j}{k}} \SM{\transp{(S; T)}(A; B)}\\
&= \max_{S \in \W{i}{k}} \max_{T\in\W{j}{k}} \SM{\transp{S}A + \transp{T}B}\\
&\le \max_{S \in \W{i}{k}} \max_{T\in\W{j}{k}} (\SM{\transp{S}A} + \SM{\transp{T}B})\\
&\le \max_{S \in \W{i}{k}} \SM{\transp{S}A} + \max_{T\in\W{j}{k}} \SM{\transp{T}B}\\
&= L_k(A) + L_k(B).
\end{align*}
\qed

\noindent Note that $L_k(A) \le L_k(A; B)$ does not hold in general. For example, let us have
\begin{equation}
A = \left(\begin{array}{rr}1&1\\1&-1\end{array}\right)
\end{equation}
and 
\begin{equation}
B = \left(\begin{array}{rr}-1&0\end{array}\right).
\end{equation}
Then by explicit calculation we obtain
\begin{equation}
4 \! = \! L_2(A) > L_2(A; B) \! = \! 3.
\end{equation}

Finally, it is shown that $L_k$ relates to the cut norm $C$, a matrix norm introduced by Frieze and Kannan in Ref.~\cite{FK97} (see also \cite{BCLSV08} for several applications in graph theory). This norm is defined as follows:
\begin{equation} 
C(M)=\max \sum_{x=1}^{n}\sum_{y=1}^{m} M_{x,y}a_xb_y,
\label{CM}
\end{equation}
where the maximum is taken over all $a_x, b_y\in\{0,1\}$. Note the similarity in the definition with the $L(M)$ norm~(\ref{LMdef}) which is equivalent to (\ref{LDef}). It has been shown that $C(M)$ is related to $L(M)$ as follows~\cite{FK97,AN04}: 
\begin{equation}
C(M)\le L(M) \le 4C(M).
\label{CMto4CM}
\end{equation}
Using the above relation~(\ref{CMto4CM}) along with Lemma~\ref{LkkL}, we find that
\begin{equation}
C(M)\le L_k(M) \le 4kC(M),
\label{CMto4kCM}
\end{equation}
and for the special case of $L_2$ we have the following lower and upper bounds: 
\begin{equation}
C(M)\le L_2(M) \le 8C(M).
\label{CMto8CM}
\end{equation}

{\it Generalization of the $L_k$ norm.}---Below we generalize the norm $L_k(M)$ to $F_M$, which extension will prove to be a key property in the Branch-and-Bound~\cite{BB} implementation of the $L_k$ algorithm. To do so, first we define the following function
\begin{definition}
\begin{equation}
\begin{aligned} &F_M : \W{i}{k} \rightarrow \mathbb R \\
&F_M(P) = \max_{W \in \W{n-i}{k}} \SM{\transp{(P; W)}M} 
\end{aligned}
\label{KDef}
\end{equation}
where \, $i = (0, 1, 2, \dots, n$)\, and\, $M \in \mathcal M_{n,m}$.
\end{definition}

In other words, \(F_M(P)\) is the maximum of $\SM{\transp{W}M}$ where $W \in\W{k}{n}$ and the prefix of $W$ is $P$. $F_M$ can be considered as a generalization of $L_k(M)$. The following lemma introduces a key property which is made use of in the Branch-and-Bound method. 
\begin{lemma}
\begin{equation}
F_{A; B}(P)  \le  \SM{\transp{P}M} + L_k(B)
\label{FfL2}
\end{equation}
\end{lemma}

{\noindent\it Proof.}
\begin{align*}
F_{A; B}(P)&= \max_{W \in \W{n-i}{k}} \SM{\transp{(P; W)}(A; B)}\\
&= \max_{W \in \W{n-i}{k}} \SM{(\transp{P}A + \transp{W}B)}\\
&\le \max_{W \in \W{n-i}{k}} (\SM{\transp{P}A} + \SM{\transp{W}B})\\
&= \SM{\transp{P}A} + \max_{W \in \W{n-i}{k}} \SM{\transp{W}B}\\
&= \SM{\transp{P}A} + L_k(B).
\end{align*}
\qed

\noindent Let us now give the following definition further generalizing \(F_M(P)\):
\begin{definition}
\begin{equation}
f_M(P)(c) = \max(F_M(P), c)
\label{KPrimeDef}
\end{equation}
\end{definition}
\noindent The computation of $f_M$ can be optimized such that for big enough $c$ values $f_M(P)(c)$ returns $c$ without computing $F_M(P)$. This is expressed by the following lemma.
\begin{lemma}
\begin{align}
&f_M(P)(c)\nonumber\\ 
&= \left\{\begin{array}{l}\max(\SM{\transp{P}M}, c) \quad\quad\quad\quad\quad\quad\;\; \text{if } P \in \W{n}{k}, \\
c \quad \text{if } \SM{\transp{P}A} + L_k(B) \le c, \quad\;\;\; A; B = M, \\
(f_M(P; \Id{k}_0) \circ \dots \circ f_M(P; \Id{k}_k))(c) \;\;\, \text{otherwise}\end{array}\right.
\label{KPrimeOptRec}
\end{align}
\end{lemma}
\noindent The proof given below is split into three cases.
\noindent Case 1: if $P \in \W{n}{k}$, then
\begin{align*}
&f_M(P)(c)\\
&= \max\left(\max_{W \in \W{n-n}{k}} \SM{\transp{(P; W)}M}, c\right) \quad \text{by}\; (\ref{KDef})\, \text{and}\, (\ref{KPrimeDef})\\
&= \max(\SM{\transp{(P; ())}M}, c)\\
&= \max(\SM{\transp{P}M}, c).
\end{align*}

\noindent Case 2:
\begin{align*}
\SM{\transp{P}A} &+ L_k(B) \le c\\
&\Rightarrow F_M(P) \le c \quad \text{by}\; (\ref{FfL2})\\
&\Rightarrow \max(F_M(P), c) = c\\
&\Rightarrow f_M(P)(c) = c \quad \text{by}\; (\ref{KPrimeDef})
\end{align*}

\noindent Case 3:
\begin{align*}
&f_M(P)(c)\\
&= \max\left(\max_{W \in \W{n-i}{k}} \SM{\transp{(P; W)}M}, c\right) \quad \text{by} \; (\ref{KDef}, \ref{KPrimeDef})\\
&= \max\left(\max_{S\in\W1k, W\in\W{n-i-1}{k}} \SM{\transp{(P; S; W)}M}, c\right)\\
&= \max\left(\max_j \max_{W \in \W{n-i-1}{k}} \SM{\transp{(P; \Id{k}_j; W)}M}, c\right)\\
&= \max(\max_j F_M(P;\Id{k}_j), c)\\
&= \max(F_M(P;\Id{k}_0), \max(F_M(P;\Id{k}_1), \max(..., c)))\\
&= (f_M(P;\Id{k}_0) \circ \dots \circ f_M(P;\Id{k}_k))(c).
\end{align*}
\qed

\noindent Our last lemma in this subsection reads
\begin{lemma}
\begin{equation}
L_k(M) = f_M(())(0)
\label{L2Alg}
\end{equation}
\end{lemma}
and it can be proved as follows:
\begin{align*}
&f_M(())(0)\\
&= \max\left(\max_{W \in \W{n}{k}} \SM{\transp{((); W)}M}, 0\right) \quad \text{by}\, (\ref{KDef}, \ref{KPrimeDef})\\
&= \max_{W \in \W{n}{k}} \SM{\transp{W}M}\\
&= L_k(M)
\end{align*}
\qed

\subsection{Programming tips for the efficient implementation of the \texorpdfstring{$L_2$}{L2} and \texorpdfstring{$L_k$}{Lk} codes}
\label{sec:L2tips}
In this subsection, we give programming tips for the branch-and-bound~\cite{BB} implementation of the exact computation of $L_k(M)$ for any $k\ge 2$. For $k=2$ and $k=3$ our algorithms are even faster than the $L_k$ solver for general $k$ due to specialization which we detail below. First, we remind the reader of the notation defined in Sec.~\ref{sec:L2proofs}. The Haskell code can be downloaded from Github~\cite{L2code}. Instructions installing and using the code (including parallel execution and using guessed results) can also be found there.

{\it Branch-and-bound calculation of $L_k$.}---The norm $L_k(M)$ for $k\ge 2$ can be calculated using the following definition and the following lemma. 

\begin{definition}
\noindent For all $M \in \Mat{n}{m}, \, 0\le i\le n$ let
\begin{align}
&f_M : \W{i}{k} \rightarrow (\mathbb R \rightarrow \mathbb R)\nonumber \\
&f_M(P)(c) \\&= \left\{\begin{array}{l}
\max(\SM{\transp{P}M}, c) \quad\quad\quad\quad\quad\quad\quad \text{if } P \in \W{n}{k},\nonumber \\
c \quad \text{if } \SM{\transp{P}A} + L_k(B) \le c, \quad\quad\; A; B = M,\nonumber \\
(f_M(P; \Id{k}_0) \circ \dots \circ f_M(P; \Id{k}_k))(c) \;\;\,\, \text{otherwise.}\end{array}\right.
\end{align}
\label{KPrimeOptRecO}
\end{definition}
\noindent The function $f_M$ recursively calls itself with larger and larger $P$ prefixes until the prefix size reaches $n$. The middle case is a conditional exit from the recursion, which speeds up the computation crucially.
\begin{lemma}
\begin{equation}
L_k(M) = f_M(())(0).
\label{L2AlgO}
\end{equation}
\end{lemma}

{\it Reducing cost by sharing sub-calculations.}---In the definition~\ref{KPrimeOptRecO}, the most expensive calculations are $L_k(B)$, $\SM{\transp{P}M}$ and $\SM{\transp{P}A}$. We show how to reduce the cost of these calculations. The cost of \(L_k(B)\) can be reduced by memoizing the previously computed \(L_k\) values in a table.

If \(M = (v_1; v_2; v_3; \dots; v_n)\) then \(L_k(M)\) depends on
\(L_k(v_i; v_{i+1}; \dots; v_n)\), where \(i = 2, 3, 4, \dots, n\). Note
that \(L_k(v_i; v_{i+1}; \dots; v_n)\) itself depends on
\(L_k(v_j; v_{j+1}; \dots; v_n)\), where \(j = i+1, i+2, \dots, n\). If
we take into account all dependencies, the correct order of calculating \(L_k\)
values is \(L_k(v_n)\), \(L_k(v_{n-1}, v_n)\),
\(L_k(v_{n-2}, v_{n-1}, v_n)\), \dots, \(L_k(v_2, v_3, \dots, v_n)\).

There is an option to skip the \(\SM{\transp{P}A} + L_k(B) \le c\) test for large
\(B\) matrices. This means that \(L_k(B)\) should not be calculated, and the trade-off is that we miss opportunities for exiting recursion. In our
experience, skipping the test for
\(B \in \Mat{k}{m}, k \ge (3n/4)\) results in about $2\times$ speedup.

The cost of calculating \(\SM{\transp{P}A}\)
is $O(km)$ if $A\in\Mat{k}{m}$.
Note that
\begin{equation}
\transp{(P; Q)}(A; B) = \transp{P}A + \transp{Q}B.
\label{share}
\end{equation}
$\transp{P}A$ is already computed by the time when $\transp{(P;\Id{k}_i)}(A;v)$
is needed, so the cost of $\SM{\transp{P}A}$ can be reduced to
\(O(m)\) by (\ref{share}). The cost of $\transp{P}M$ can be reduced in the same way.
This implies a considerable speedup; for example,
for \(M \in \mathcal M_{70, 70}\) the calculation of \(L_2(M)\) can be made nearly 70 times faster by this optimization.

The cost of $\SM{\transp{P}A}$ can be further reduced by caching the previously calculated
Manhattan norms of the rows of the matrix $\transp{P}A$.

{\it Reducing cost by symmetries.}---For all $S\in\Perm{k}$ permutation matrices
\begin{equation}
\SM{\transp{W}M} = \SM{\transp{(WS)}M}.
\label{symm}
\end{equation}
\noindent The cost of $L_2$ can be halved by (\ref{symm}) as follows.
Let 
\begin{equation}
S = \left(\begin{array}{rr}0&1\\1&0\end{array}\right)\in\Perm{2}.
\nonumber
\end{equation}
\noindent From $\SM{\transp{(\Id{2}_2; W)}M} = \SM{\transp{(\Id{2}_1; WS)}M}$
it follows that $f_M(\Id{2}_2, c) = f_M(\Id{2}_1, c)$.
This means that we can skip the calculation of $f_M(\Id{2}_2, c)$ for all $c$, thus
$L_2(M) = f_M((\Id{2}_1), 0)$, i.e., we start the calculation with a non-empty prefix which saves work.

Harnessing (\ref{symm}) in the general $L_k$ case is a bit more complex.
First we define the set of canonical prefixes. A prefix $P = \Id{k}_{i_1}; \Id{k}_{i_2}; \dots \Id{k}_{i_j}$
is canonical if the first occurrences of the numbers in the indices $i_1, i_2, \dots, i_j$ is the sequence
$1, 2, 3, \dots$. For example, the prefix
 $\Id{k}_1; \Id{k}_2; \Id{k}_1; \Id{k}_3$ is canonical
but $\Id{k}_1; \Id{k}_3; \Id{k}_1; \Id{k}_2$ is non-canonical.
For each prefix $P$, there exists a permutation $S$ such that $PS$ is canonical, so that,
$f_M(P, c) = f_M(PS, c)$, which means that it is enough to examine only the canonical prefixes to compute $L_k$.

{\it Parallel and concurrent execution.}---For parallel execution one can use the following equation:
\begin{equation}
f_M(P)(c) = \max_{i\in\{1,\dots,k\}} f_M(P;\Id{k}_i)(c)
\label{fPar}
\end{equation}
We used Eq.~(\ref{fPar}) for $P\in\W{i}{k}, i<d$, where $d$
is a ``{parallel depth}'' for fine-tuning the execution for different architectures.
Higher depth is better for more cores.

Parallel execution may miss opportunities of exiting recursion because there is no communication between threads about the best known $L_k$ values at a certain point of time. Therefore we implemented concurrent execution where threads share the best known $L_k$ values.

{\it Reducing cost by guessed $L_k$ values.}---Optionally, the computation can be sped up by providing a guessed $L_k(M)$ value by the user.
This value will be used instead of $0$ in Eq.~(\ref{L2AlgO}).
The guessed value may be lower than $L_k(M)$.
Higher guessed values are better, unless the guessed value is higher than $L_k(M)$, in which case
$f_M$ returns the guessed value. We compared the result of $f_M$ with the witness $W$ of the maximal $\SM{\transp{W}M}$ value, to be able to detect whether the guessed value was too high or not.

\subsection{\texorpdfstring{$L$}{L}-norm and \texorpdfstring{$L_2$}{L2}-norm are the same for a special family of matrices \texorpdfstring{$M'$}{Mv}}
\label{sec:LL2Mp}

We relate $L(M')$ to $L_2(M')$, where $M'$ is given by the following matrix
\begin{align}
M'=\begin{pmatrix}
	&M \\
 -&M 
\end{pmatrix},
\label{Mv}
\end{align}
where $M$ is a matrix of size $n\times m$ with arbitrary real entries. Note that $M'$ has size $2n\times m$ and $M'$ has rows such that $M'_x=M_x$ and $M'_{x+m}=-M_{x}$ for all $x=1,\ldots,m$. Then the following lemma holds.

\begin{lemma}
$L_2(M')=L(M')=2L(M)$ for any matrix $M'$ of the form (\ref{Mv}), where $L_2$ is the $L_2$ norm given by the definition~(\ref{L2Mv3}) and $L$ is the local bound given by~(\ref{LDef}). Note that $L_k$ is defined by~(\ref{L2Def}), where the case $k=2$ corresponds to the definition of $L_2$ in (\ref{L2Mv3}).  
\end{lemma}

{\noindent\it Proof.} 
We fix a matrix $M$ of dimension $n\times m$ which specifies $M'$ by the virtue of (\ref{Mv}). Let $a_x\in\{-1,1\}$, $x=1,\ldots,n$ and $b_y\in\{-1,1\}$, $y=1,\ldots,m$ be the optimal vectors giving $L(M)$ in (\ref{LDef}). Note that these values are not unique in general, different optimal configurations may exist, however, we choose one such optimal vectors $a_x$ and $b_y$. We then choose $a_{x+n}=-a_x$ for $x=1,\ldots,n$, and $b_y^+=b_y^-=b_y$ for $y=1,\ldots,m$. With these values, we obtain the lower bound $L_2(M')\ge 2L(M)$ on $L_2(M')$ in (\ref{L2M}).
Now we show the upper bound $L_2(M')\le 2L(M)$, which implies $L_2(M')=2L(M)$.

As a contradiction of the lemma, assume that $L_2(M')>2L(M)$. Then, not all $a_x$ vectors corresponding to the $L_2(M')$ value have the property $a_{x+m}=-a_x$ for each $x$. That is, there exists at least one $x$, call it $x'$, for which  $a_x'=a_{x'+n}$. Suppose that there is one such an $x'$ (the proof for multiple $x'$ indices for which  $a_x'=a_{x'+n}$ is very similar). 
Then in the formula~(\ref{L2Mv3}) for $L_2(M')$ the two rows $x'$ and $x'+n$ in question will appear within the same norm (either in the first or second norm, depending on whether $a_x'=a_{x'+n}$ takes the value $+1$ or $-1$). However, in both cases they cancel each other from the norm in question. As a result, two rows of $M'$ in (\ref{Mv}) are eliminated, one from the matrix $M$ and one from the matrix $-M$. However, any matrix $\pm M$ from which one row has been eliminated cannot have a local bound greater than $L(M)$. The same applies to a matrix $\pm M$ from which we have removed several rows. Therefore, $L_2(M')>2L(M)$ cannot be true either. Thus we arrived at a contradiction. $\qed$

\subsection{Adapting the Gisin-Gisin model to the PM scenario}
\label{sec:gisinmodel}

We now adapt the LHV model of Ref.~\cite{GG} which exploits the finite efficiency of the detectors to reproduce the quantum correlations of the singlet state exactly. We show that the LHV model in Ref.~\cite{GG} can be adapted to the PM communication scenario to produce the expectation value: 
\begin{equation}
\label{EGG}
E(\vec a,\vec b)=P(b=+1|\vec a,\vec b)-P(b=-1|\vec a,\vec b)=\frac{\vec a\cdot\vec b + 1}{2},
\end{equation}
where $\vec a\in S^2$ denotes the preparation Bloch vector and $\vec b\in S^2$ denotes the measurement Bloch vector. First we show that the outcomes $b=\pm 1$ giving the expectation value 
\begin{equation}
\label{EGG2}
E(\vec a,\vec b)=P(b=+1|\vec a,\vec b)-P(b=-1|\vec a,\vec b)=\vec a\cdot\vec b
\end{equation}
can be obtained with probability $1/2$ and $b=0$ outcome with probability $1/2$. Then by coarse-graining the above distribution by grouping $b=0$ outcome with $b=+1$, we obtain the expectation value~(\ref{EGG}).

The classical model, using one bit of classical communication from Alice to Bob, is as follows. 

{\noindent\it Protocol:} Alice and Bob share a classical variable, which is in the form of a unit vector $\vec\lambda$, chosen uniformly at random from the unit sphere $S^2$.
\begin{itemize}
\item {\it (Alice)} Alice sends a binary message $c=\sign(\vec a\cdot\vec\lambda)$ to Bob. That is, $c=+1$ if $\vec a\cdot\vec\lambda\le 0$ and $c=-1$ if $\vec a\cdot\vec\lambda> 0$.
\item {\it (Bob)} Bob outputs $b=\sign(c\vec b\cdot\vec\lambda)$ with probability $|\vec b\cdot\vec\lambda|$ (corresponding to the detection event $b=\pm 1$) and Bob outputs $b=0$ with probability $1-|\vec b\cdot\vec\lambda|$ (corresponding to the non-detection event).
\end{itemize}  

Our claim is as follows. The above protocol yields the correlations $E(\vec a,\vec b)=\vec a\cdot\vec b$, that is, it reproduces the correlations in Eq.~(\ref{EGG2}) with probability $1/2$ and returns $b=0$ in the other cases.

{\it Proof.}---We need to calculate the expectation value $E(\vec a,\vec b)=P(b=+1|\vec a,\vec b)-P(b=-1|\vec a,\vec b)$ which according to the above protocol in the detection events $b=\pm 1$ is given by~\cite{GG}
\begin{equation}
E(\vec a,\vec b)=\int_{S^2}d\vec\lambda q(\vec\lambda|b=\pm 1)\sign(\vec a\cdot\vec\lambda)\sign(\vec b\cdot\vec\lambda),
\label{Eab2}
\end{equation}
where $q(\vec\lambda|b=\pm 1)$ is the conditional density probability distribution of choosing $\vec\lambda$ given a detection event (either output $b=+1$ or $b=-1$). This function can be calculated from 
\begin{equation}
q(\vec\lambda|b=\pm 1)=\frac{q(\vec\lambda \text{ and } b=\pm 1)}{p(b=\pm 1)},
\label{qlambdab}
\end{equation} 
where the detection efficiency is $\eta=p(b=\pm 1)$ and the probability of detection failure is $1-\eta=p(b=0)$. The value of $\eta$ is given by 
\begin{equation}
\eta=p(b=\pm 1)=\int_{S^2}\frac{d\vec\lambda}{4\pi}|\vec b\cdot\lambda|=\frac{1}{2}, 
\end{equation}
as stated and the protocol gives the density probability distribution $q(\lambda \text{ and } b=\pm 1) = (\vec b\cdot\vec\lambda)/(4\pi)$. Inserting these values into (\ref{qlambdab}) gives $q(\lambda|b=\pm 1)=(\vec b\cdot\vec\lambda)/(2\pi)$, which in turn is inserted into (\ref{Eab2}) to obtain the integral~\cite{GG}:
\begin{equation}
E(\vec a,\vec b)=\frac{1}{2\pi}\int_{S^2}d\vec\lambda(\vec b\cdot\vec\lambda)\sign(\vec a\cdot\vec\lambda).
\label{Eab3}
\end{equation}
The above integral can be calculated using spherical symmetries. In particular, one can choose w.l.o.g. the vectors
\begin{align*} 
\vec a &= (0,0,1)\\
\vec b &=(\sin\alpha,0,\cos\alpha)
\end{align*}
as in Ref.~\cite{GG}, and then obtain
\begin{equation}
E(\vec a,\vec b)=\cos\alpha=\vec a\cdot\vec b
\label{Eab4}
\end{equation}
with probability $1/2$, which we wanted to prove. \qed

\subsection{The modified Gilbert algorithm adapted to the PM scenario}
\label{sec:gibi}

\noindent For a given $\eta\in\left[1/2,1\right]$ and correlation matrix $E(\eta)$ defined by (\ref{Exyeta}), the algorithm yields the following matrix $M$ satisfying 
\begin{equation}
\label{hyper2}
\sum_{x=1}^n\sum_{y=1}^m M_{xy}E_{xy}(\eta)>L_2(M).
\end{equation}

\noindent \textbf{Algorithm:} 

{\rm Input}: The number of preparations $n$ and the number of measurement settings $m$ that define the setup. The unit vectors $\{\vec a_x\}_{x=1}^n$ (i.e., the Bloch vectors of Alice's prepared states) and $\{\vec{b_y}\}_{y=1}^m$ (i.e., the Bloch vectors of Bob's projective rank-1 measurements). The $(n\times m)$-dimensional matrix $E(\eta)$ given by the entries $E_{xy}(\eta)$ in (\ref{Exyeta}). The values of $\epsilon$ and $i_{\text{max}}$ that define the stopping criteria.

{\rm Output}: The matrix $M$ of size $n\times m$.

\begin{enumerate}
\item Set $i=0$ and set $E^{(i)}$ the $n\times m$ zero matrix. 
\item Given a matrix $E^{(i)}$ and the matrix $E(\eta)$, run a heuristic oracle that maximizes the overlap $\sum_{xy}(E_{xy}(\eta)-E_{xy}^{(i)})E_{xy}^{\text{det}}$ over all deterministic one-bit correlations $E_{xy}^{\text{det}}$ in (\ref{Exybit}). 
The description of this heuristic (see-saw) oracle is given in Sec.~\ref{sec:L2seesaw}. Denote the point $E_{xy}^{\text{det}}$ returned by the oracle by $E_{xy}^{\text{det,i}}$. 
\item Find the convex combination $E^{(i+1)}$ of $E^{(i)}$ and $E_{xy}^{\text{det,i}}$ that minimizes the distance $\sqrt{\sum_{xy}\left(E_{xy}(\eta)-E_{xy}^{(i)}\right)^2}$. Let us denote this distance by $\text{dist}(i)$.
\item Let $i=i+1$ and go to Step 2 until $\text{dist}(i)\le\epsilon$ or $i=i_{\text{max}}$. 
\item Return the matrix $M$ with coordinates $M_{xy}=E_{xy}(\eta)-E_{xy}^{(i)}$.
\end{enumerate}

Note that $\text{dist}(i)$ is a decreasing function of $i$. Since maximizing the overlap of $\sum_{xy}(E_{xy}(\eta)-E_{xy}^{(i)})E_{xy}^{\text{det}}$ over all deterministic one-bit correlations vectors is an NP-hard problem, in Step 2 we use a heuristic method to do it, which we describe in Sec.~\ref{sec:L2seesaw}. On the other hand, the description of an exact branch-and-bound type algorithm can be found in Sec.~\ref{sec:L2tips}. We use the exact method, which is generally more time-consuming than the see-saw method to check that the output matrix $M$ satisfies the condition~(\ref{hyper2}) with the chosen parameter $\eta$. If this is true, then it implies the lower bound $K_{\text{D}}\ge (1/\eta)$, as proved in Sec.~\ref{sec:uplow}. It should also be noted that the branch-and-bound-type algorithm is much faster than the brute force algorithm (the implemented algorithm using parallelism can be found in \cite{L2code}). On a multi-core desktop computer, it can solve problems in range $n=m=70$ in a day, while the brute force algorithm is limited to about $n=m=40$ settings.

\subsection{Parameters and implementation of Gilbert algorithm}
\label{sec:techgibi}
Here we specify the explicit parameters that are used to obtain the lower bound $K_{\text{D}}\le 1.5682$. On the three-dimensional unit sphere, we choose the vectors $\{\vec a_x\}_{x=1}^n$ and $\{\vec b_y\}_{y=1}^m$  to be equal to each other, $\vec v_i = \vec a_i=\vec b_i$ for $i=1,\ldots,n$, where $n=m=70$. The 70 unit vectors chosen define the optimal packing configuration in the Grassmannian space which can be downloaded from Neil Sloane's database~\cite{Sloane2}. The advantage of this type of packing is that the points and their antipodal points are located as far apart as possible on the three-dimensional unit sphere. 

We implemented the modified Gilbert algorithm (of section~\ref{sec:gibi}) in Matlab with and without a memory buffer (see more details on the memory buffer in Ref.~\cite{Gilbert}). In the case of using memory buffer, the step 3 is modified in the algorithm so that instead of calculating the convex combination of the points $E^{(i)}$ and $E_{xy}^{\text{det,i}}$ (see section~\ref{sec:gibi}), we compute the convex combination of $E^{(i)}$ and the points $E_{xy}^{\text{det,i-j}}$, $j=0,\ldots,m-1$, where $m$ is the size of the memory buffer. In our explicit computations, we use a buffer size $m=40$ and a stopping condition of $k=2\times 10^5$ with $\eta=0.665$. Details on the performance of this modification can be found in Ref.~\cite{Gilbert}. In step 2 of the Gilbert algorithm, the oracle uses the see-saw heuristic described in Sec.~\ref{sec:L2seesaw} to obtain a good (typically tight) lower bound to $L_2(M)$. On the other hand, we used the branch-and-bound-type algorithm described in Sec.~\ref{sec:L2tips} to calculate $L_2(M)$ exactly for integer $M$. The algorithm was implemented in Haskell. See the GitHub site~\cite{L2code} for the downloadable version.

The Matlab file \verb|eta_70.m|, which can also be downloaded from GitHub~\cite{L2code} (located in the subdirectory \verb|L2_eta_70|) gives detailed results on the input parameters. In particular, it gives the unit vectors $\vec a_i=\vec b_i=\vec v_i$, the lower bound $\sum_{xy}M_{xy}\vec a_x\cdot\vec b_y=\sum_{xy}M_{xy}\vec v_x\cdot\vec v_y$ to $q(M)$  and the value $L_2(M)$. The input matrix $M$ is placed in subdirectory \verb|L2_eta_70| under the name \verb|W70i.txt|. 
The running time of the Gilbert algorithm (in Sec.~\ref{sec:gibi}) implemented in Matlab was about one week. Note, however, that most of the computation time was spent on the oracle (the see-saw part) described in Sec.~\ref{sec:L2seesaw}. On the other hand, the Haskell code to compute the exact $L_2(M)$ value of the $70\times 70$ witness matrix $M$ took about 8 hours to run on a HP Z8 workstation using 56 physical cores. The memory usage of the computation was negligible. 

The Matlab \verb|eta_70.m| routine defines the $70\times 70$ matrix $M$, and gives the $\vec v_i:=\vec a_i=\vec b_i$ the unit vectors from Sloane's database~\cite{Sloane2} for all $i=1,\ldots,70$. Note that $M$ is integer (by multiplying the output $M$ matrix in the Gilbert algorithm by $1000$ and truncating the non-integer part). This calculation yields 
$S(M)=\sum_{x,y}M_{x,y}=194369$ and $Q(M)=\sum_{x,y}M_{x,y}\vec a_x\cdot\vec b_y\simeq5.3672235\times10^5$. On the other hand, the branch-and-bound-type Haskell code~\cite{L2code} gives the exact value $L_2(M)=412667$, which is matched by the see-saw search (in Sec.~\ref{sec:L2seesaw}). From these numbers we then obtain
\begin{equation}
K_{\text{D}}\ge\frac{Q(M)-S(M)}{L_2(M)-S(M)}>\frac{342353}{218298}=1.5682+\varepsilon
\end{equation}
and $1/K_{\text{D}}=0.6377-\varepsilon'$ is the upper bound to the critical detection efficiency $\eta_{\text{crit}}$, where $\varepsilon$ and $\varepsilon'$ are small positive numbers.



\subsection{Lower bound to \texorpdfstring{$L_2(M)$}{L2M} using the see-saw iterative algorithm}
\label{sec:L2seesaw}

Below we give an iterative algorithm based on see-saw heuristics to compute $L_2(M)$. This algorithm forms the oracle part of step 2 of the Gilbert algorithm, which is described in Sec.~\ref{sec:gibi}.

\noindent \textbf{Algorithm:} 

{\rm Input}: Integer matrix $M$ of size $n\times m$.

{\rm Output}: Lower bound $l_2(M)$ to $L_2(M)$ defined by formula~(\ref{L2M}).
\begin{enumerate}
 \item Let $l_2 = 0$.
 \item Choose random assignments $a_x = \pm 1$: That is, $a_x$ are (random) elements of a vector $a$ of size $n$. Its elements are binary having value +1 or -1 only.
 \item Set $b^{+} = \text{sgn}(aM)$, where \text{sgn} denotes the (modified) sign function: $\text{sgn}(x)=+1$ if $x\ge 0$ and $-1$ otherwise. Let us transpose $b^{+}$.
 \item Set $b^{-} = \text{sgn}(aM)$. Let us transpose $b^{-}$.
 \item Form the column vector $s^+=Mb^+$ of size $n$.
 \item Form the column vector $s^-=Mb^-$ of size $n$. 
 \item Form the column vector $s=\max(s^+,s^-)$ of size $n$. That is, $s_x=\max(s_x^+,s_x^-)$ for all $x=1,\ldots,n$.
 \item Form the $\pm 1$-valued column vector $a$ as follows: Let $a_x=+1$ if $s_x^+\ge s_x^-$, otherwise let $a_x=-1$ for all $x=1,\ldots,n$.
 \item Let $l_2=\sum_{x=1}^{n}s_x$.  
 \item{With the new vector $a$, return to point 3. Repeat the algorithm until two values of $l_2$ are equal in two consecutive iterations.}
\end{enumerate}
Note that at each iteration step, objective value $l_2(M)$ is guaranteed not to decrease. Therefore, the output of the algorithm is a heuristic lower bound on the exact value of $L_2(M)$. 

\section{Discussion}

We have tested the quantumness of two-dimensional systems in the prepare-and-measure (PM) scenario, with $n$ preparations and $m$ binary-outcome measurement settings, where $n$ and $m$ fall well into the range of 70. In the one-qubit PM scenario, a two-level system is transmitted from the sender to the receiver. In this setup, a real $n\times m$ matrix $M$ defines the coefficients of a linear witness. We denote by $L_2(M)$ the exact value of the one-bit bound associated with matrix $M$. We found efficient numerical algorithms for computing $L_2(M)$. If this bound is exceeded, we can detect both the quantumness of the prepared qubits and the quantumness (i.e. incompatibility) of the measurements. 

We introduced new constants $K_{\text{M}}$ and $K_{\text D}$ which are related to the Grothendieck constant of order 3. Our large-scale tools are crucial for the efficient bounding of $L_2(M)$ and hence for bounding of the constants $K_{\text{M}}$ and $K_{\text D}$. We further relate these new constants to the white noise resistance of the prepared qubits and the critical detection efficiency of the measurements performed.

For large $M$ matrices, we have given two algorithms for computing $L_2(M)$: a simple iterative see-saw-type algorithm and a branch-and-bound-type algorithm. The former is a heuristic algorithm that usually gives a tight lower bound on $L_2(M)$. However, sometimes it fails to find the exact value of $L_2(M)$. This happens more and more often as the size of the matrix $M$ gets larger and larger. 
In contrast, the latter branch-and-bound-type algorithm gives the exact value of $L_2(M)$ and can be used to compute $L_2(M)$ for matrix sizes as large as $70\times 70$. As an application of the algorithms, we established the bounds $1.5682\le K_{\text D}\le 2$ on the new constant and an upper bound of $\eta_{\text{crit}}\le 0.6377$ on the critical detection efficiency of qubit measurements in the PM scenario.

\section*{Data availability} The dataset (the $70\times 70$ matrix $M$) used to prove the lower bound (\ref{KD_LU}) on $K_D$ is available in Github~\cite{L2code}.

\section*{Code availability} The Haskell and MATLAB codes used to prove the lower bound (\ref{KD_LU}) on $K_D$ is available in Github~\cite{L2code}.

\section*{Acknowledgements} We thank M\'aty\'as Barczy, Emmanuel Zambrini Cruzeiro and Armin Tavakoli for valuable discussions. We are particularly indebted to M\'aty\'as Barczy for pointers to the literature regarding the cut norm. T. V. acknowledges the support of the EU (QuantERA eDICT) and the National Research, Development and Innovation Office NKFIH (No. 2019-2.1.7-ERA-NET-2020-00003).


\end{document}